# New Physics in High Landau Levels

J.P. Eisenstein[1], M.P. Lilly[1], K.B. Cooper[1], L.N. Pfeiffer[2] and K.W. West[2]

[1]*California Institute of Technology, Pasadena, CA 91125*
[2]*Bell Laboratories, Lucent Technologies, Murray Hill, NJ 07974*

## Abstract

Recent magneto-transport experiments on ultra-high mobility 2D electron systems in GaAs/AlGaAs heterostructures have revealed the existence of whole new classes of correlated many-electron states in highly excited Landau levels. These new states, which appear only at extremely low temperatures, are distinctly different from the familiar fractional quantum Hall liquids of the lowest Landau level. Prominent among the recent findings are the discoveries of giant anisotropies in the resistivity near half filling of the third and higher Landau levels and the observation of re-entrant integer quantum Hall states in the flanks of these same levels. This contribution will survey the present status of this emerging field.

## Introduction

The discovery and study of the fractional quantum Hall effect (FQHE) has dominated the discussion of strong correlation physics in two-dimensional electron systems for over 15 years[1]. The existence of a diverse family of incompressible quantum liquids as well as novel "composite fermion" compressible states makes the early predictions[2] of charge density wave ground states a striking example of the failure of Hartree-Fock theory. On the other hand, essentially all of this physics concerns the behavior of many-electron systems in the lowest (N=0) Landau level. Until recently, remarkably little research has been done on the problem of electron-electron correlations in the excited Landau levels.

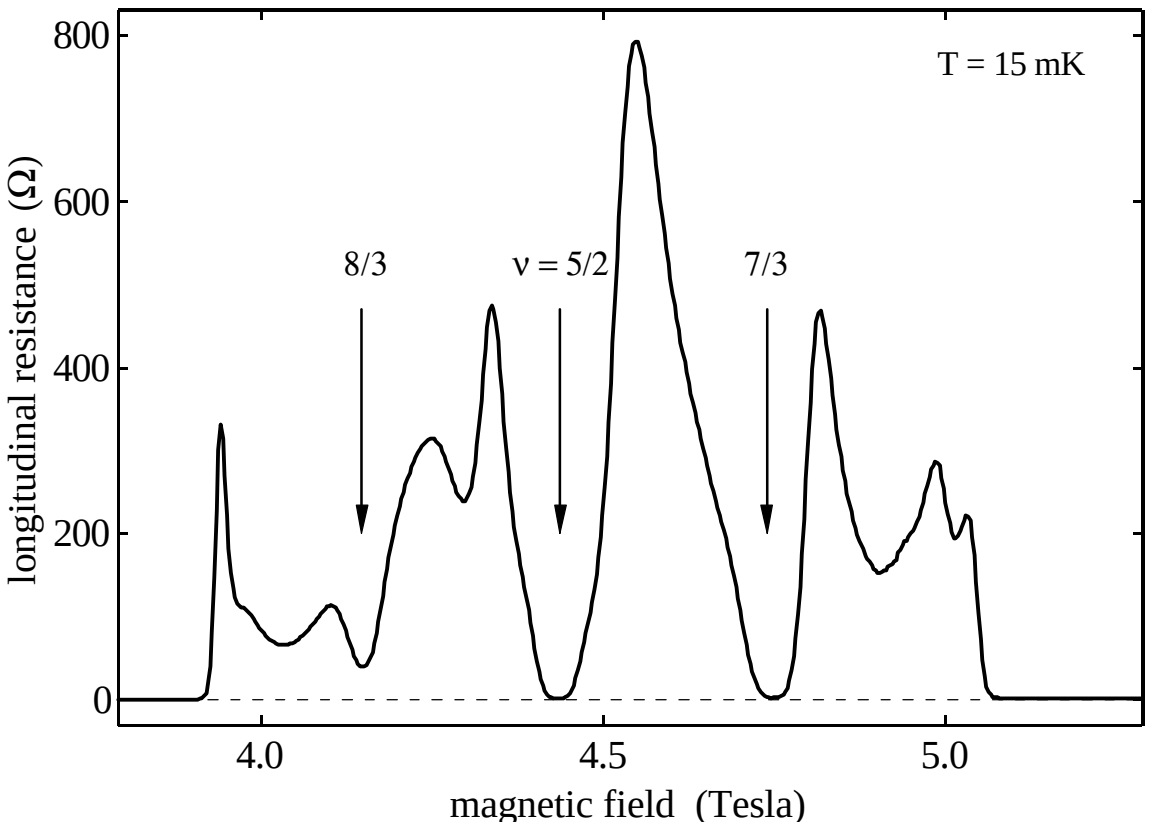

**Fig. 1.** Resistivity data from the N=1 Landau level taken at T=15mK. The 5/2, 7/3, and 8/3 FQHE states are indicated.

That the situation in these higher levels is not merely a replica of lowest Landau level physics has actually been evident for quite some time. In 1987 the first FQHE state having an even-denominator was discovered[3] at filling factor $\nu=5/2$ in the N=1 second Landau level (LL). Since then, in spite of great improvements in sample quality, the only other even-denominator FQHE known is the ν=7/2 state[4]. This state, which is also in the N=1 LL, is presumably the particle-hole conjugate of the 5/2 state. Figure 1 shows resistivity data in the vicinity of ν=5/2 taken at T=15mK using a high quality GaAs/AlGaAs heterostructure having a mobility of $11\times10^6 cm^2/Vs$. Although the 5/2 state is quite well developed in this sample, it is accompanied by *only two* odd-denominator FQHE states, at ν=7/3 and 8/3. This near-absence of the FQHE in the N=1 level contrasts sharply with the plethora of states seen in the N=0 level.

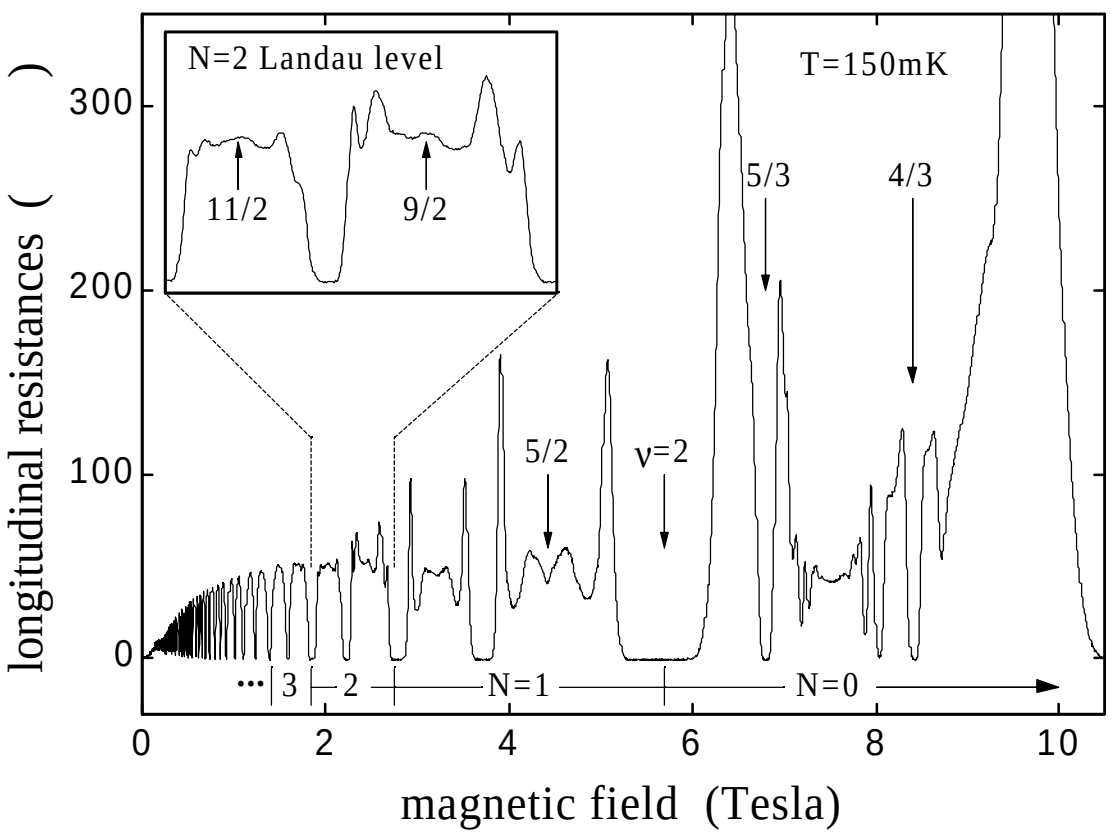

**Fig. 2.** Overview of the longitudinal resistivity at T=150mK in an $11x10^6 cm^2/Vs$ mobility 2D electron gas. Data from the N=2 Landau level is expanded in the inset.

In the *third* and higher LLs (N≥2 and ν≥4) the situation changes again. Little attention has been paid to this low magnetic field regime owing to the expectation that subtle correlation effects would be overwhelmed by disorder. Nonetheless, as early as 1988 there were hints from transport experiments that interesting things might be going on[5]. Figure 2 shows longitudinal resistivity data obtained at T=150mK from the same sample as used for Fig.1. Above B=2.7T, the Fermi level is in either the N=1 or N=0 LL. Familiar odd-denominator fractional quantum Hall states are seen in the N=0 level and the emerging ν=5/2 and 7/2 states in the N=1 level are also evident. In the inset the data from the N=2 LL (filling factors between 6>ν>4) are magnified and substantial structure in the resistivity is apparent. This structure is not consistent with a disorder-driven transition between integer quantized Hall states and is instead clear evidence for electron correlations. To determine the nature of these correlations obviously requires lower temperatures.

The remainder of this paper will survey, in outline form, the prominent findings of recent transport measurements on high mobility 2D electron gases in at high Landau level filling factors. While the first conclusive reports of dramatic transport anomalies were those of the Caltech/Bell group[6,7], essentially identical findings have been reported by the Utah/Columbia/Princeton/NHMFL/Bell collaboration[8,9]. Most recently, related findings have also been reported for 2D *hole* systems by Shayegan and Manoharan[10].

## II. Anisotropic Transport

Figure 3 shows the temperature development of the resistance in the vicinity of ν=9/2. Two traces are shown at each temperature: these correspond to two different current flow configurations in the 5x5mm square sample. In both cases the current is driven between two ohmic contacts placed at the midpoints of opposite sides of the sample and the voltage is detected using two contacts at the corners along one side of the current flow axis. For the dashed trace the average current flow direction is along the <1-10> crystallographic direction; the resistance measured thusly is denoted $R_{xx}$. For the solid trace the current is along the orthogonal <110> direction and the resistance is $R_{yy}$.

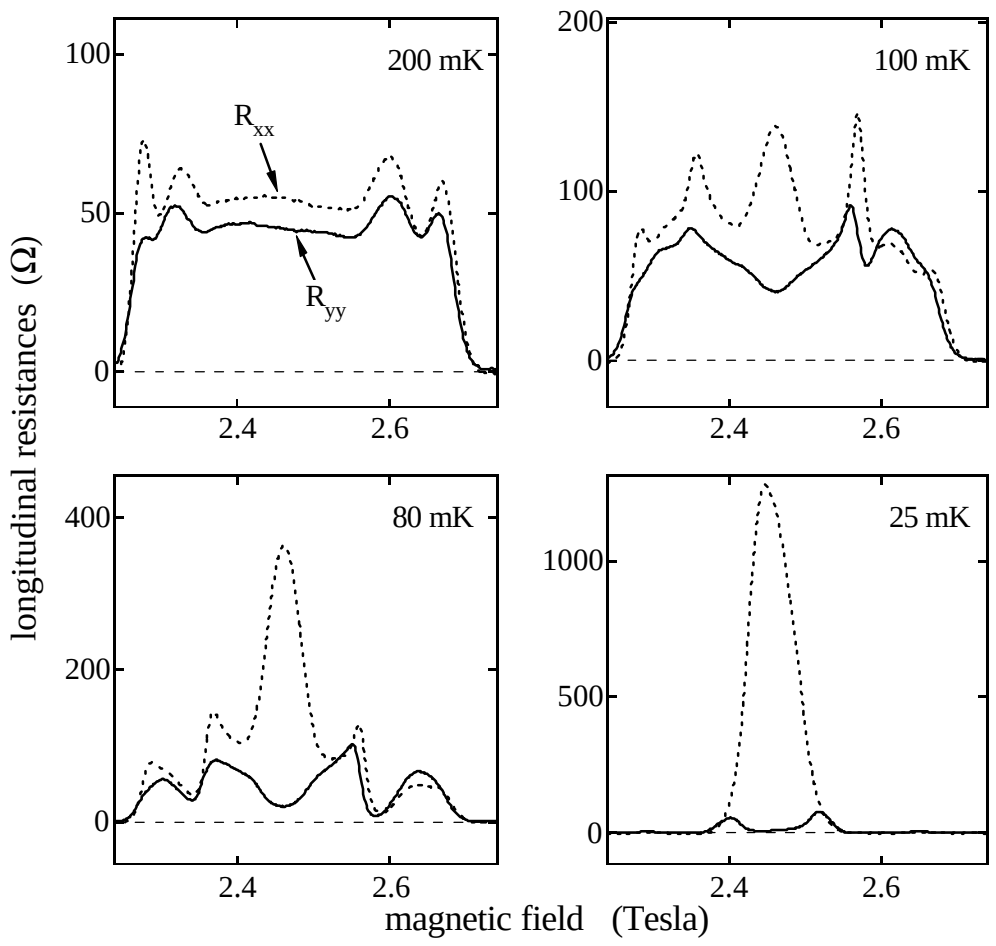

**Fig. 3**. Temperature development of resistance features around ν=9/2.

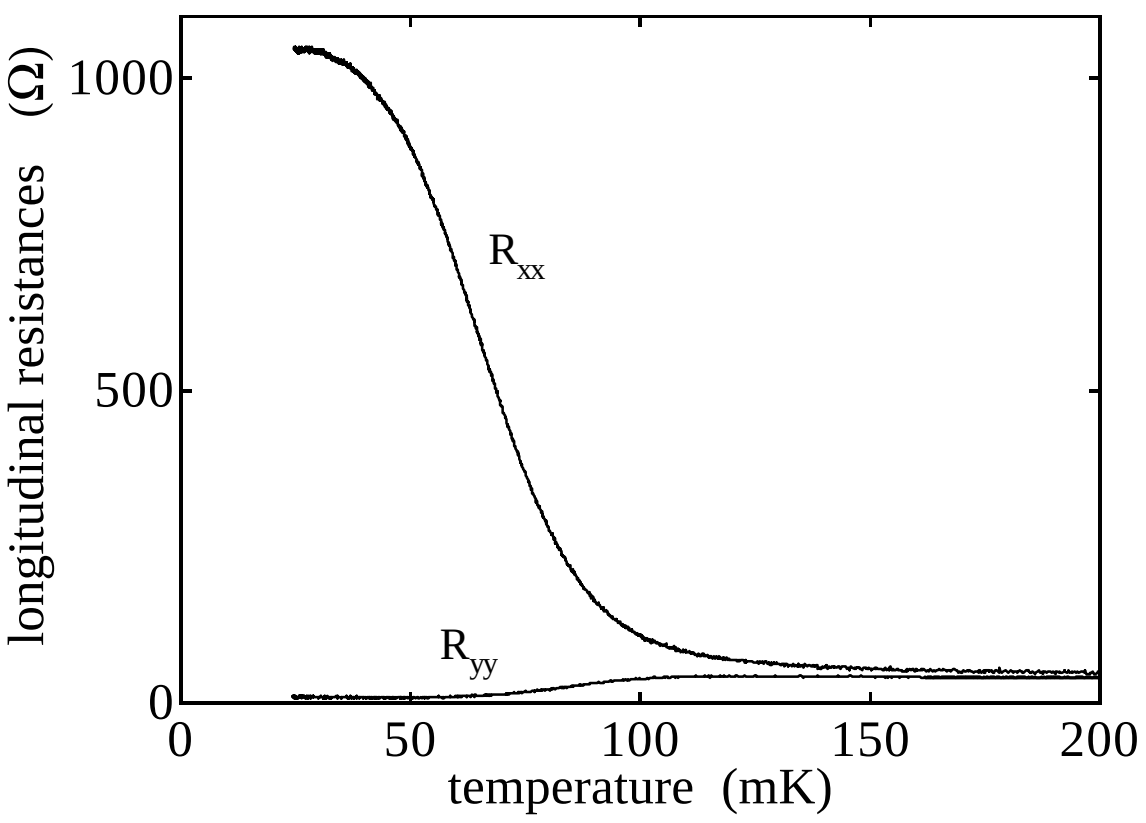

**Fig. 4**. Resistances at exactly ν=9/2.

(Note that the $R_{yy}$ trace has been multiplied by 0.6.) At high temperatures the two traces are very similar but below about 150mK a clear difference develops around half-filling. A maximum develops in $R_{xx}$ while a minimum forms in $R_{yy}$. Note that the peak in $R_{xx}$ does not narrow as the temperature is reduced. By T=25mK the difference between the two measurements has become enormous (about 100x) and dwarfs the factor of 0.6 originally applied to facilitate the comparison. The development of this giant anisotropy in the resistance is perhaps the most striking aspect of transport in high Landau levels in clean 2D electron systems. Figure 4 shows the rapid way in which the anisotropy at ν=9/2 develops as the temperature is reduced below 100mK. Notice that below about 40-50mK both $R_{xx}$ and $R_{yy}$ have apparently saturated at finite values. This saturation seems to be genuine and not simply a heating effect.

Figure 5 shows an overview of the resistances $R_{xx}$ and $R_{yy}$ at T=25mK. The insets clarify the two measurement configurations. It is clear that the anisotropy seen around ν=9/2 is also present at ν=11/2, 13/2, 15/2, and with decreasing strength at still higher half-odd integer fillings. Most importantly, however, the anisotropy is *absent* at ν=7/2 and 5/2 in the N=1 LL and (though not shown) in the N=0 lowest LL. This simple observation is perhaps the clearest indication that correlation phenomena in the N≥2 levels are quite different from those in the lower two Landau levels.

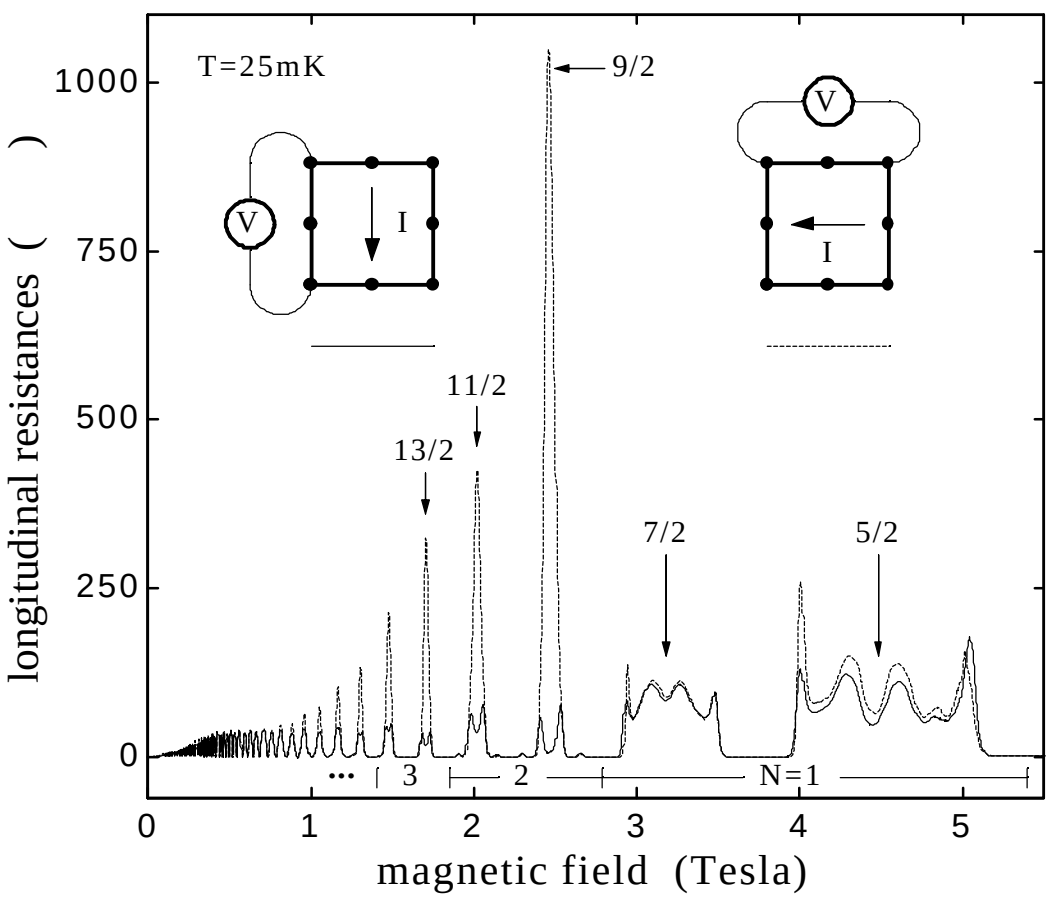

**Fig. 5**. Overview of transport anisotropy in high Landau levels.

## III. Re-entrant Integers

Away from half-filling of the N≥2 levels, the resistance again becomes essentially isotropic. Figure 6 shows data from a sample having low temperature mobility of $\mu=15.6\text{x}10^6\text{cm}^2/\text{Vs}$. For this sample, the observed resistances $R_{xx}$ and $R_{yy}$ differ by about a *factor or 3500* at $\nu$=9/2[11]. In contrast, in the flanks of the Landau level the two resistances return to approximate equality (at the factor-of-two level). Furthermore, both resistances drop to essentially zero[6,8] near *B*=1.9 and 2.1T ($\nu \approx$4.25 and 4.75), suggesting the existence of new fractional quantized Hall states. Figure 6 reveals that while the Hall resistance *is* quantized in these regions, it is at the value of the nearby *integer* quantum Hall states. (The apparent plateaus in $\rho_{xy}$ elsewhere in Fig. 6 are not stable and disappear at very low temperatures. We believe they are artifacts arising from mixing of the longitudinal and Hall resistances.) These observations therefore suggest that at these filling factors the electrons in the N=2 Landau level have assumed an *insulating* configuration. The fact that these new phases are well separated from the conventional integer QHE states suggests that the origin of the insulating behavior is not simply single particle localization; electron correlations are clearly playing an essential role. We emphasize that this re-entrant integer QHE is not found in the N=1 or N=0 Landau levels but is apparently present in several LLs with N≥2. This effect represents another of the surprising new results from high Landau levels.

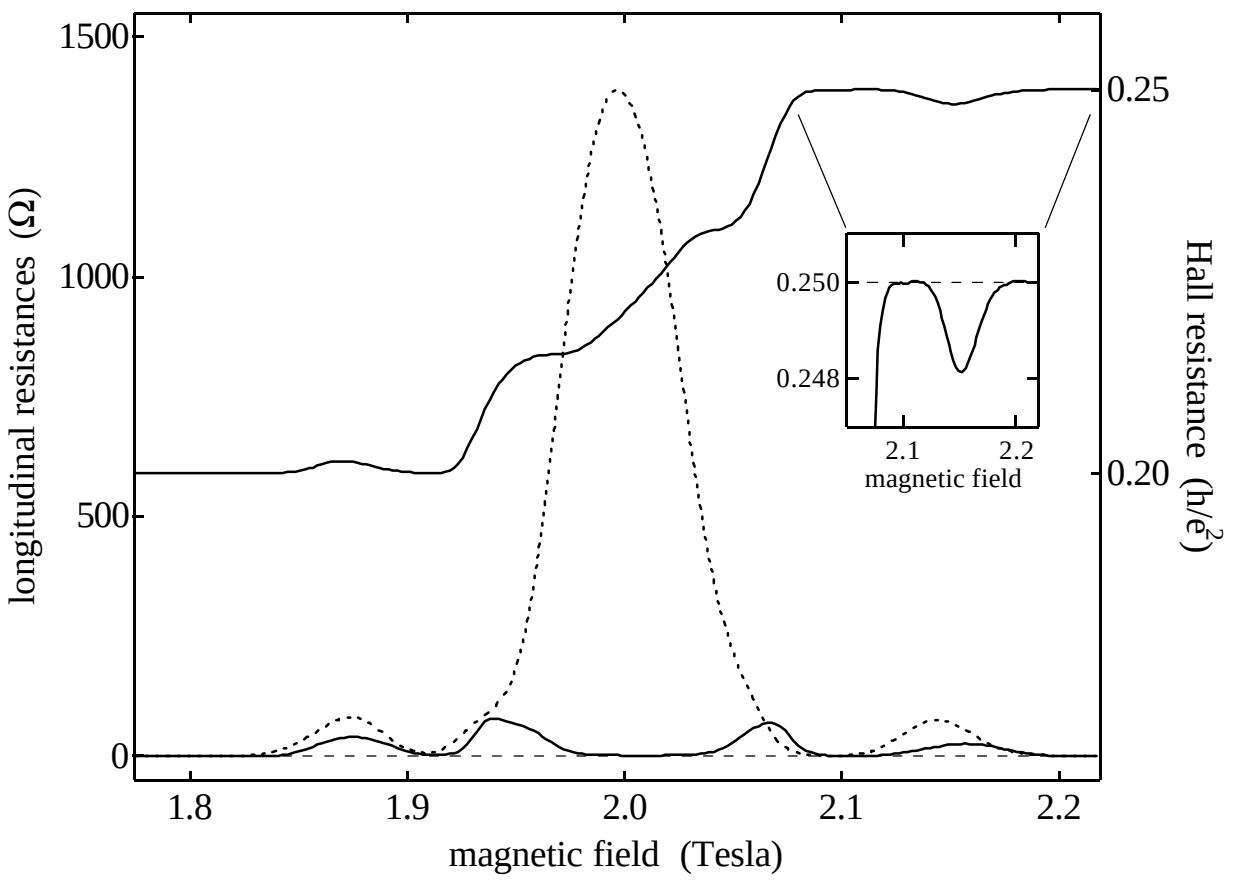

**Fig. 6**. Hall and longitudinal resistance in the range 4<$\nu$<5. Inset shows the re-entrant $\nu$=4 integer QHE.

## IV. Stripes and Bubbles?

At the qualitative level, these observations overlap strongly with the recent theoretical suggestions of charge density wave (CDW) states in high Landau levels. The early work of Koulakov, Fogler and Shklovskii[12] and Moessner and Chalker[13] has been supplemented recently by several additional contributions[14-18]. Unlike the situation in the lowest Landau level, these theoretical works conclude that Hartree-Fock CDW states are in fact good starting points for understanding the many-electron ground state in the N≥2 LLs. Near half-filling of the LLs, the CDW is predicted to be a unidirectional stripe phase in which the filling factor $\nu_N$ of the uppermost LL jumps between $\nu_N$=0 and 1. The wavelength of this stripe pattern is expected to be a few times the classical cyclotron radius, or about 1000-1500Å under typical experimental conditions. It is certainly plausible that if such a phase were coherently oriented over macroscopic dimensions it could produce a highly anisotropic resistivity. At the same time, the theories also predict that in the flanks of the LL there is a transition from the stripe phase to “bubble” phases and eventually a Wigner crystal. Such phases would likely be isotropic and probably pinned by disorder. This too is broadly consistent with the experimental observations.

Beyond these qualitative comparisons however, there are many important unanswered questions. How and why the stripes are pinned coherently across the sample is only the most obvious one. At a deeper level, the basic nature of transport in stripe or bubble phases is not understood. Should the resistances $R_{xx}$ and $R_{yy}$ measured at half-filling saturate at low temperatures? Should the width of the resistance peak remain finite as T→0? Can the existence of the re-entrant integer QHE state in the flanks of the LL be incorporated into a scenario of bubble or Wigner crystal phases? What about interaction effects beyond Hartree-Fock? Are the stripes themselves stable against a Peierls-type instability[15]? Do quantum fluctuations produce liquid crystalline behavior?[14] Obviously, much remains to be understood.

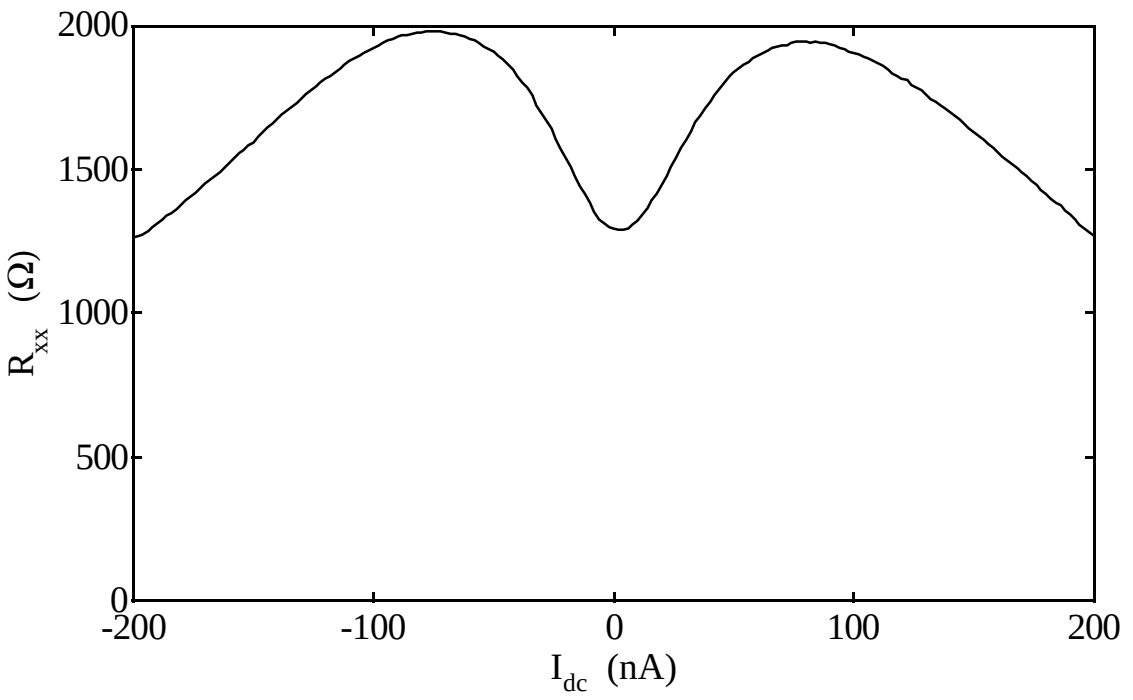

**Fig. 7**. Non-linearity at T=25mK in the resistance $R_{xx}$ at $\nu$=9/2.

## V. Non-linearities

Less heralded, but nonetheless intriguing, are the substantial non-linearities of the resistivity observed around half filling of high LLs by Lilly, *et al.*[6]. Figure 7 shows the effect on the peak resistance $R_{xx}$ of a dc current added to the small ac current used for the measurement. (The resistance $R_{xx}$ is really $dV_{xx}/dI$). For small dc currents, the resistance $R_{xx}$ *rises*. Since $R_{xx}$ falls rapidly with increasing temperature, the observed non-linearity at small dc current is inconsistent with simple electron heating. As the figure shows, larger dc current do indeed drive $R_{xx}$ down and this could well be a heating effect. It is interesting to consider what the enhancement of $R_{xx}$ at small dc current implies for the conductivity of the system. Inversion of the resistivity tensor of an anisotropic system yields $\sigma_{yy} = \rho_{xx}/(\rho_{xx}\rho_{yy}+\rho_{xy}^2)$. Since the Hall resistance $\rho_{xy}$ is much larger (about 6kΩ at $\nu$=9/2) than both $R_{xx}$ and $R_{yy}$, we have approximately $\sigma_{yy} \approx \rho_{xx}/\rho_{xy}^2$. Thus, a large *resistance* $R_{xx}$ in the <1-10> direction is equivalent to a large *conductance* in the <110> direction. Hence, within the stripe model, the enhancement of $R_{xx}$ at small dc current implies enhancement of the conductivity along the stripes. The data in Fig. 7 show that this enhancement turns on continuously rather than exhibiting a sharp threshold as expected for the depinning of a charge density wave. On the other hand, Fertig[15] has recently suggested that the non-linearity we observe is consistent with a model of a quantum melted modulated stripe phase. In Fertig's model the Hall electric field along the stripes produced by the dc current increases the population of charged soliton-antisoliton pairs and thereby increases the conductance.

## VI. Tilted Fields

To further examine the anisotropic phases near half filling, tilted magnetic field studies have been performed[7,9]. The in-plane magnetic field $B_{||}$ created by tilting the sample couples to the system through the spin Zeeman energy and the finite thickness of the electronic wavefunction in the direction perpendicular to the 2D plane. Beyond this, however, the in-plane field also breaks the rotational symmetry of the system. It is important to see how this symmetry breaking interacts with the internal symmetry breaking field which macroscopically orients the intrinsic transport anisotropy of half-filled high Landau levels in the first place.

Figure 8 shows the $B_{||}$-dependence of the resistances $R_{xx}$ (dashed curves) and $R_{yy}$ (solid curves) at T=50mK at $\nu$=9/2 and 11/2 in the N=2 LL. In the left two panels $B_{||}$ is directed along the <110> crystal direction while in the right two it is along <1-10>. Recall that (at $B_{\|}=0$) transport along the <110> direction yields the minimum in $R_{yy}$ while transport along <1-10> gives the tall peak in $R_{xx}$. It is clear from the figure that theses resistances are quite sensitive to both the magnitude and direction of the in-plane magnetic field.

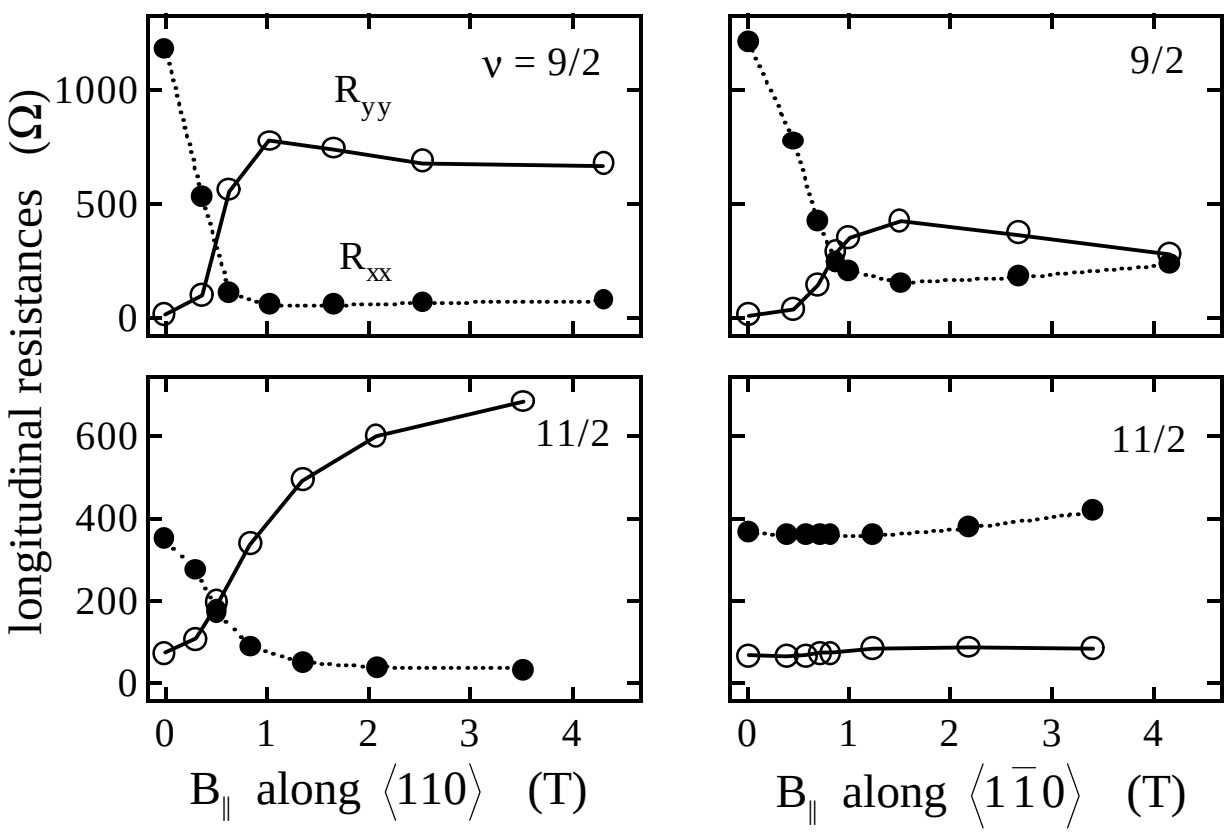

**Fig. 8**. Left two panels: Interchange of "hard" and "easy" transport directions induced by an in-plane magnetic field along <110>. Right panels: Effect of an in-plane field along <1-10>.

With $B_{\|}$ oriented along <110> the large anisotropy ($R_{xx}$>>$R_{yy}$) at $B_{\|}=0$ gives way to the opposite condition at large in-plane field. This remarkable "interchange effect" is also seen at $\nu=13/2$ and 15/2 in the N=3 LL and in several higher half-filled levels as well. It appears that the in-plane magnetic field rotates the anisotropy axes by 90° in the 2D plane. Interestingly, the interchange seems to always to occur at $B_{\|}\approx 0.5$T. If one adopts a stripe model, and assumes that transport transverse to the stripes produces higher resistance than transport along them, then these data suggest the stripes want to orient themselves perpendicular to $B_{\|}$. Recent theoretical calculations[17,18] have shown that the preferred orientation of stripes is sensitive to numerous quantitative details. Nonetheless, the most realistic of such calculations[18] are in agreement with the experiment. These calculations also provide a useful estimate of the built-in symmetry breaking field: 10mK/electron is typical.

The data for $B_{\|}$ along <1-10> (right-hand panels of Fig. 8) shows that the situation is not quite so simple. The above discussion leads to the prediction that if the stripes are already perpendicular to $B_{\|}$, the in-plane field should have little effect. At $\nu=11/2$ this is indeed the case: there is essentially no dependence of $R_{xx}$ and $R_{yy}$ on $B_{\|}$. At $\nu=9/2$, however, there is a substantial effect[19]. The resistances become roughly equal as the in-plane field increases. This is a particularly puzzling result, especially since $\nu=9/2$ and 11/2 are in the *same* N=2 orbital LL. Similar dependences on the spin sub-level are seen in essentially every aspect of these new anisotropic correlated phases in high Landau levels. We remind the reader, however, that breakdowns of particle-hole symmetry are seen in the lowest and first excited LLs as well.

## VII. The N=1 Landau Level

The even-denominator fractional quantum Hall state at $\nu=5/2$ in the N=1 LL has been one of the enduring mysteries of the 2D electron field. Early tilted field studies[20] showing the collapse of the 5/2 state were widely interpreted as indicating the existence substantial spin reversal in the ground state, but that conclusion has been brought into question, especially by recent numerical studies[21]. The early tilt experiments, however, missed an important point. In addition to suppressing the 5/2 (and 7/2) FQHE state, the tilting leaves the transport in the N=1 LL highly anisotropic. Figure 9 shows data at zero and large tilt angle demonstrating both of these effects. For the data shown, $B_{||}$ is along <110>, but very similar results are found for $B_{||}$ along <1-10>. As in the case of the N≥2 Landau levels, at large tilt angle higher resistance is found for transport along the direction of the in-plane magnetic field[22].

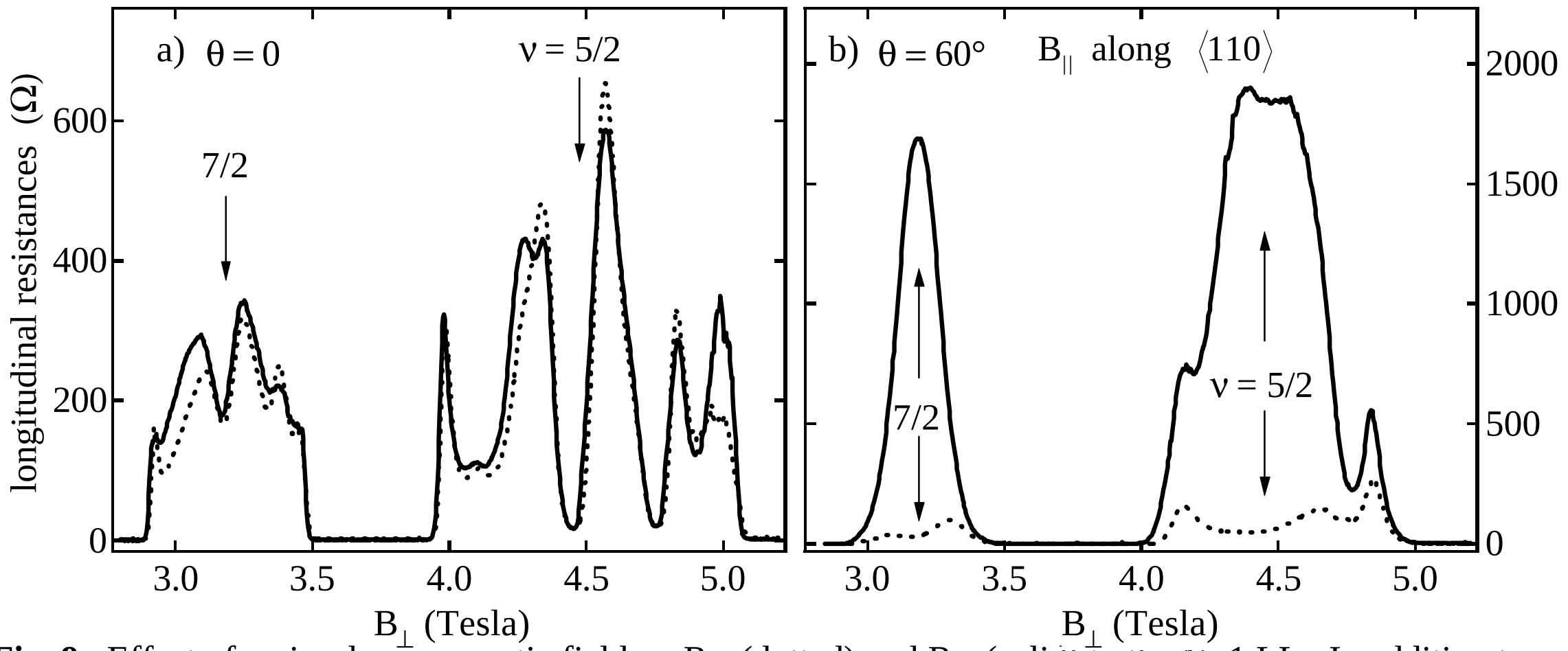

**Fig. 9**. Effect of an in-plane magnetic field on $R_{xx}$ (dotted) and $R_{yy}$ (solid) in the N=1 LL. In addition to destroying the ν=5/2 and 7/2 FQHE states, the in-plane field leaves the transport highly anisotropic.

These results do not explain the mechanism for the destruction of the 5/2 FQHE state itself. However, they do suggest that the 5/2 state is close in energy to an anisotropic correlated phase similar to those found at half filling of the higher Landau levels. This is an interesting finding since one might have expected the 5/2 FQHE state to be replaced, at large tilt angle, by a composite fermion liquid phase like that observed at ν=1/2 in the N=0 lowest Landau level. These qualitative ideas have found support in recent numerical calculations.[23]

## VIII. Summary and Outlook

The recent transport experiments on ultra-clean 2D electron systems at high Landau level filling have yielded a number of fascinating discoveries. Highly anisotropic correlated electron phases near half filling and intriguing new insulators in the wings of the Landau levels have been found. These new phenomena appear in several adjacent Landau levels, provided N≥2. This suggests a generic mechanism and heightens the overall significance of the results. The absence of the fractional quantum Hall effect (so far, at least) in these same Landau levels is also remarkable. In the N=1 Landau level new tilted field experiments show that the 2D electron system is somehow at a boundary between fractional quantum physics and the new, possibly striped, world of high Landau levels.

Obviously, much remains to be done. The bulk transport measurements discussed here offer no direct way to determine whether a charge density modulation on the scale of 1000Å exists or not. Other techniques are needed to probe the system on this length scale. Surface acoustic waves and inelastic light scattering are obvious candidates. Application of scanned probes, though still in the dark ages from the standpoint of QHE physics, offers another multi-year funding opportunity. What is clear at this point is that a new class of 2D electron correlated states has been uncovered. Rumors of the death of the 2D electron field have, once again, been exaggerated.